# Modeling water infiltration into soil under fractional wettability conditions

Simone Di Prima [a,*], Ryan D. Stewart [b], Majdi R. Abou Najm [c], Deniz Yilmaz [d], Alessandro Comegna [a] and Laurent Lassabatere [e]

[a] Department of Agricultural, Forestry, Food and Environmental Sciences (SAFE), University of Basilicata, 85100 Potenza, Italy.
[b] School of Plant and Environmental Sciences, Virginia Polytechnic Institute and State University, Blacksburg, VA, United State.
[c] Department of Land, Air and Water Resources, University of California, Davis, CA 95616, United States.
[d] Univ. Grenoble Alpes, CNRS, IRD, Grenoble INP, IGE, 38000 Grenoble, France.
[e] Université de Lyon; UMR5023 Ecologie des Hydrosystèmes Naturels et Anthropisés, CNRS, ENTPE, Université Lyon 1, Vaulx-en-Velin, France.
* Corresponding Author. E-mail: simone.diprima@unibas.it

## Highlights

- We propose a new infiltration model for fractional wettable conditions
- The soil surface is modeled as an ensemble of two distinct fractions
- We validated the new model against both analytical and field data.

## Abstract

The heterogeneous distribution of water-repellent materials at the soil surface causes a phenomenon known as fractional wettability. This condition frequently triggers destabilization of the wetting front during water infiltration, resulting in the formation of fingered bypass flow. However, few analytical tools exist to understand and model this behavior. Moreover, existing infiltration models fail to fit certain infiltration curves that exist in experimental data. For these reasons, we introduce a novel infiltration model to simulate water infiltration under fractional wettable conditions. We conceptualize the soil surface as a composite of two distinct portions: a water-repellent fraction, where hydrophobic effects impede water infiltration, and a wettable fraction, where capillarity and gravity are the dominant forces controlling the process. The new model was validated using a dataset comprising infiltration data from 60 field measurements. Additionally, validation was performed using 660 analytically generated infiltration curves from six synthetic soils with varying textures. This innovative approach enabled us to account for the combined influence of these two fractions and to enhance the interpretation of infiltration curves with mixed shapes, which other common methods are unable to reproduce.

## 1. Introduction

Water repellency is a property that commonly affects the soil surface layer. It results from hydrophobic coatings on soil particles that originate from organic matter (Doerr et al., 2000). The most significant effect of soil water repellency is a reduction in infiltration rates. However, although dry soils can initially display strong water repellency, this effect diminishes during the wetting of the soil (DeBano, 2000). Additionally, the infiltration of water into water-repellent soils often triggers instability of the wetting front, leading to fingered bypass flow, a phenomenon frequently observed when analyzing wetting patterns in soil profiles after infiltration tests (e.g., Dekker and Ritsema, 2000; Doerr et al., 2006; Ganz et al., 2013; Lichner et al., 2018; Robinson et al., 2010).

Modeling water infiltration under these circumstances remains a significant challenge. Indirect strategies have been employed in numerical modeling to assess the impact of soil water repellency on the estimation of soil hydraulic parameters (e.g., Diamantopoulos et al., 2013; Filipović et al., 2018; Nguyen et al., 1999; Nieber et al., 2000). A related strategy is to assess hysteresis in soil water retention, under the assumption that soil water repellency will primarily influence the wetting direction (Bauters et al., 1998). Modeling software programs, such as HYDRUS 2D/3D (Šimůnek et al., 2008), have also been used to assess soil water repellency based on hysteresis. For example, Ganz et al. (2014) used the hysteresis model of Lenhard and Parker (1992), which is implemented in HYDRUS, to account indirectly for the effect of soil water repellency. While they successfully simulated the conical finger geometry, they concluded that dynamic approaches are necessary due to the highly variable nature of water repellency in both space and time.

At the field scale, fingered bypass flow is linked to the uneven distribution of hydrophobic material on the soil surface, as discussed by Ritsema and Dekker (1994). In such scenarios, the soil surface can be conceptualized as a mosaic, featuring alternating hydrophobic and wettable regions. During a rainstorm, water falling on initially dry hydrophobic regions is impeded from infiltrating, whereas water infiltrates more readily into wettable regions. Infiltration rates into wettable areas are governed by the hydraulic characteristics of the soil matrix, including permeability, sorptivity, and initial moisture content. In hydrophobic regions, infiltration is also influenced by the specific properties of the hydrophobic material present. This configuration is referred to as fractional wettability (FW). Beatty and Smith (2013) delineated a fractional wettable system as one in which a portion or fraction comprises soil particles with water-wettable surfaces, while another fraction comprises soil particles characterized by water-repellent surfaces.

Macroscopically, infiltration in fully wettable soils generates concave-shaped cumulative infiltration curves, whereas infiltration in dry hydrophobic soils produces convex-shaped cumulative infiltration curves. The behavior of water-repellent soils may fall between these two extremes and may exhibit a transition from concave to convex shapes or vice versa. These mixed shapes are influenced by various factors, including soil moisture content and the proportion of the soil exhibiting either hydrophilic or hydrophobic characteristics (Chen et al., 2020). According to Pachepsky and Karahan (2022), up to five distinct shape types can be found in water-repellent soils, including: convex (K), convex-to-linear (E) and concave-to-convex (L) shapes in case of soils having high organic carbon content, and linear (H) and regular concave (A) shapes in case of low water repellent conditions (uppercase letters were used in their work as shape type codes). Soil water repellency is also a highly dynamic property that changes over time, typically decreasing as the soil water content increases (Dekker and Ritsema, 1994). This variable wettability not only makes static measurements of water repellency inadequate but also hinders the accurate reproduction of convex-shaped cumulative infiltration curves when using conventional infiltration models (Angulo-Jaramillo et al., 2019).

Numerical simulations using the inverse modeling option in HYDRUS have been conducted to model water infiltration in water-repellent soils (e.g., Wang et al., 2018; Yang et al., 2021). However, these simulations failed to reproduce convex or mixed infiltration shapes. This limitation arises because the Richards equation is inadequate for hydrophobic soils due to the dynamic nature of soil water repellency (Diamantopoulos and Durner, 2013). Recently, new analytical formulations have been proposed, leading to significant improvements in modeling water infiltration in water-repellent soils (e.g., Abou Najm et al., 2021; Berli and Shillito, 2023; Di Prima et al., 2021; Hammecker et al., 2022; Moret-Fernández and Latorre, 2023; Yilmaz et al., 2022). For instance, Moret-Fernández and Latorre (2023) tested a new physically-based model to describe infiltration curves with complex shapes. These authors tested the model in the laboratory through one-dimensional infiltration tests, demonstrating the model's ability to fit convex-to-linear cumulative infiltration curves. Abou Najm et al. (2021) proposed implementing an empirical exponential scaling factor $(1 - e^{-\alpha_{WR} t})$ into infiltration models, which accounts for the time-varying water repellency at the soil surface. Di Prima et al. (2021) utilized this correction factor to modify the three-dimensional (3D) two-terms (2T) explicit transient infiltration model created using a second-order Taylor expansion of the quasi-exact implicit (QEI) model developed by Haverkamp et al. (1994). They validated the new model against both synthetic and real soils. Similarly, Yilmaz et al. (2022) proposed modifications to the three-terms (3T) formulation by Rahmati et al. (2019) using a third-order Taylor expansion of the same QEI model. The 3T model underwent testing on the same set of synthetic soils used by Di Prima et al. (2021), demonstrating better performance in many scenarios compared to the 2T version. However, both models exhibited poor fitting in cases of fractional wettability occurrences, particularly when the shape of the cumulative infiltration curve was neither consistently concave nor convex. Indeed, while the incorporation of the correction factor $(1 - e^{-\alpha_{\text{WR}} t})$ has been shown to accurately reproduce convex-shaped curves, modeling other shapes still remains a challenge.

In this investigation, we propose a new infiltration model to simulate water infiltration under fractional wettable conditions. The novel approach views the soil surface as an ensemble of two distinct fractions: (1) a water-repellent (WR) fraction, where hydrophobic effects impede water infiltration, leading to an increase in the infiltration rate over time and convex-shaped cumulative infiltration curves; and (2) a wettable (W) fraction, where capillarity and gravity forces predominate, resulting in a decreasing infiltration rate over time until reaching a steady state and exhibiting concave-shaped cumulative infiltration curves. This new approach allows us to account for the combined influence that these two fractions have on the shape of cumulative infiltration curves. Curves generated using this model can exhibit different degrees of convexity, depending on the severity of the hydrophobic effect, as well as mixed forms such as concave-to-convex and convex-to-concave. The new model was validated using the infiltration dataset collected by Di Prima et al. (2021), which included data from 60 field measurements and was really needed to allow accurate fits. Additionally, validation was performed against 660 analytically generated infiltration curves from six synthetic soils with different textures and degrees of water repellency, to simulate a comprehensive set of scenarios.

## 2. Theory

We assume that water repellency exists as a discrete layer that exists at or just below the soil surface and covers some portion of the infiltration area (Figure 1). With this assumption, water infiltration into a fractional wettability soil, $I_{\text{FW}}(t)$, can be modeled by the following equation:

$$I_{\text{FW}}(t) = I_{\text{W}}(t) \times w_{\text{FW}} + I_{\text{WR}}(t) \times (1 - w_{\text{FW}}) \qquad 0 \leq w_{\text{FW}} \leq 1 \tag{1}$$

where $I_{\text{W}}(t)$ is the three dimensional (3D) cumulative infiltration into the wettable fraction, $I_{\text{WR}}(t)$ is the 3D cumulative infiltration into the water-repellent fraction, and the subscripts W and WR respectively refer to wettable and water-repellent. The parameter $w_{\text{FW}}$ represents the fraction of the total water volume infiltrating into the wettable soil. Consequently, $(1 - w_{\text{FW}})$ represents the remaining volume fraction infiltrating water through the water-repellent surface. The parameter $w_{\text{FW}}$ expresses the relative importance of the forces interacting in the fractional wettability soil. The proposed approach is schematically illustrated in Figure 1.

We can model $I_{\text{W}}(t)$ using the following transient infiltration equation for concave-shaped curves (Lassabatere et al., 2006):

$$I_{\text{W}}(t) = S\sqrt{t} + [A(1 - B)S^2 + B i_s] t \tag{2}$$

where $t(T)$ is the time elapsed since the start of the infiltration event, $S(\text{LT}^{-0.5})$ is the soil sorptivity, $i_s(\text{LT}^{-1})$ is the steady-state infiltration rate, $B$ is a coefficient that can be set equal to 0.467 for most soils with dry initial conditions (Di Prima et al., 2016), and $A$ ($\text{L}^{-1}$) is defined as follows:

$$A = \frac{\gamma}{r(\theta_s - \theta_i)} \tag{3}$$

where $r$(L) is the radius of the infiltration source, $\theta_s$(L³L⁻³) and $\theta_i$(L³L⁻³) are the saturated and initial volumetric soil water contents, and $\gamma$ is a shape parameter for geometrical correction of the infiltration front shape, which is commonly set to 0.75 (Haverkamp et al., 1994). Note that, although many alternative formulations are available in the literature (Angulo-Jaramillo et al., 2019), we chose Eq. (2) because it has been extensively tested and proven to provide accurate estimates of soil hydraulic parameters, particularly saturated soil hydraulic conductivity (Di Prima et al., 2016).

We can then model $I_{WR}(t)$ using the equation proposed by Di Prima et al. (2021) for convex-shaped curves:

$$I_{WR}(t) = S\sqrt{t} - \frac{S\sqrt{\pi}}{2\sqrt{\alpha_{WR}}} erf(\sqrt{\alpha_{WR}t}) + [A(1-B)S^2 + Bi_s]t - \frac{[A(1-B)S^2 + Bi_s](1 - e^{-\alpha_{WR}t})}{\alpha_{WR}} \tag{4}$$

where $(1 - e^{-\alpha_{WR}t})$ is an exponential scaling factor proposed by Abou Najm et al. (2021) to modify the infiltration rate for better simulations of water-repellent conditions, and in which the empirical parameter $\alpha_{WR}$ (T⁻¹) reflects the rate of water repellency attenuation during infiltration. Again, our choice is consistent since Eq. (4) results from the application of the correction factor to Eq. (2). In other words, Eqs. (2) and (4) can be seen as approximate expansions for the transient state of the QEI model without versus with water repellency.

By applying Eqs. (2) and (4), Eq. (1) becomes:

$$I_{FW}(t) = \{S\sqrt{t} + [A(1-B)S^2 + Bi_s]t\} \times w_{FW} + \\ \left\{S\sqrt{t} - \frac{S\sqrt{\pi}}{2\sqrt{\alpha_{WR}}}erf(\sqrt{\alpha_{WR}t}) + [A(1-B)S^2 + Bi_s]t - \frac{[A(1-B)S^2 + Bi_s](1 - e^{-\alpha_{WR}t})}{\alpha_{WR}}\right\} \times (1 - w_{FW}) \tag{5}.$$

We note that in the case of fully wettable conditions and concave-shaped curves, Eq. (5) aligns with Eq. (2). This condition manifests under two circumstances: the first, more intuitive scenario is when $w_{FW} = 1$ and $I_{WR}(t) = 0$. The second scenario arises when $w_{FW} = 0$, yet $I_{WR}(t)$ resembles the conventional formulation of Eq. (2) due to a large $\alpha_{WR}$ value (Di Prima et al., 2021).

Note also that, theoretically, Eq. (2) addresses to the modeling of transient state before the attainment of steady state (Lassabatere et al., 2009). Similarly, Eq. (4) was obtained by applying the correction proposed by Abou Najm et al. (2021) for water repellency to Eq. (2), meaning that this expression should also be restricted to the transient state. Consequently, the proposed model should be restricted to a time interval before steady state.

## 3. Material and methods
### 3.1. Experimental assessment

For the experimental assessment of Eq. (5) we used the same dataset collected by Di Prima et al. (2021) at the Berchidda experimental site (40°48'57.28"N, 9°17'33.09"E; Sardinia, Italy). The presence of fractional wettability at the Berchidda site was supported through observations made during the field campaign conducted by Di Prima et al. (2021). Fingered flows were noted from three cross-sectional profiles that were excavated at the conclusion of dye infiltration experiments.

The site is a Mediterranean wooded grassland system characterized by herbaceous grasslands dominated by annual species and interspersed evergreen oak trees (*Quercus suber* L. and *Quercus ilex* L.). With a mean annual rainfall of 632 mm, 70% of which occurs from October to May, and a mean annual temperature of 14.2°C, the Berchidda site aligns with the climatic conditions of the Mediterranean region. Soils sampled from the upper horizon exhibited textures ranging from sandy loam to loamy sand, and were categorized as Typic Dystroxerepts according to USDA standards.

At the Berchidda site, Di Prima et al. (2021) conducted sixty infiltration tests at randomly selected sampling points, specifically around three designated trees. Among these, thirty tests were taken below the tree canopies (ten beneath each tree), while the remaining thirty were carried out in the open grasslands (ten in the proximity of each tree within the open spaces). The measurements were performed using the automated single-ring infiltrometer introduced by Di Prima (2015). Additional details regarding the device and data processing can be accessed online at the following address: https://bestsoilhydro.net/infiltrometer/. The interpretation and subsequent categorization of the curves were based on the observed arrays of shapes and on the model parameters $w_{FW}$ and $α_{WR}$. This procedure also considered the criterion proposed by Pachepsky and Karahan (2022) for categorizing infiltration curves from the global SWIG database (Rahmati et al., 2018).

Water drop penetration time ($WDPT$) tests (Wessel, 1988), which provide valuable insights into the persistence, distribution, and variability of water repellency (Dekker and Ritsema, 2000), were conducted at eighteen sampling points. For each tree, three tests were carried out below the canopy and three additional tests were conducted in the open grassland near the tree. For each test, ten drops (0.05 mL) of distilled water were placed on the soil surface using a pipette, and the time until complete infiltration of each drop was measured. A representative $WDPT$ value for each situation was obtained by averaging the thirty $WDPT$ measurements.

### 3.2. Analytical validation

The analytical validation of Eq. (5) involved two main steps. The first step consisted of generating 660 infiltration curves for six synthetic soils using Eq. (1), ten empirical $α_{WR}$ shape factor and eleven $w_{FW}$ ratios (i.e., 6 × 10 × 11 = 660 curves). The diverse selection of soils, chosen from Carsel and Parrish (1988), represents a broad range of hydraulic behaviors and includes the following soils: sand, loamy sand, sandy loam, loam, silt loam, and silty clay loam (Table 1). The second step involved the inversion of these synthetic curves using Eq. (5) and the estimation of $S$, $α_{WR}$ and $w_{FW}$. The results were subsequently compared with the parameters used to generate the original curves.

Note that pre-existing modeling software such as HYDRUS was not used in this investigation for validating the new analytical infiltration model. Indeed, the Richards equation is not considered applicable for hydrophobic medium due to the dynamic nature of soil water repellency. The simulation of convex and mixed-shaped cumulative infiltration curves would require formulating an ad-hoc adaptation of the Richards equation that incorporates a new law or new soil hydraulic functions capable of simulating effects analogous to the empirical scaling factor proposed by Abou Najm et al. (2021) and the fractional wetting concept. To date, there does not exist any implementation of the Richards equation capable of simulating and reproducing convex or mixed cumulative infiltration shapes.

### 3.2.1. STEP 1: generating synthetic curves for six soils

To generate the synthetic curves, we modeled water infiltration into the wettable fraction, $I_w(t)$ (i.e., the first term on the right side of Eq. 1), using the model proposed by Smettem et al. (1994), which includes the QEI model developed by Haverkamp et al. (1994) and is presented as Eq. (8) in Di Prima et al. (2021). Otherwise, we modeled water infiltration into the water-repellent fraction, $I_{WR}(t)$ (i.e., the second term on the right side of Eq. 1), using the QEI model developed by Haverkamp et al. (1990) and adapted to water repellent soils by Di Prima et al (2021), which is also presented as Eq. (13) in Di Prima et al. (2021).

We modeled the curves under an initially dry condition, corresponding to a saturation degree value, of $Se = 0.1$. Subsequently, this value was converted for each soil to the equivalent initial volumetric water content, $\theta_i$ (m$^3$ m$^{-3}$), using the relationship $Se = (\theta_i - \theta_r)/(\theta_s - \theta_r)$. The sorptivity was then estimated using the $\theta_r$ values from Table 1 and the flux concentration model of Parlange (1975), which involves the integration of a function incorporating the hydraulic conductivity function (Lassabatere et al., 2023).

For the empirical parameter $\alpha_{WR}$, we considered 10 different values for each soil, adjusting $\alpha_{WR}$ between 0.04 to 10000 h$^{-1}$ depending on the soil type (Di Prima et al, 2021). This range of values was intended to encompass a diverse array of shapes, ranging from regular concave to convex (Table 1).

For the parameter $w_{FW}$, we considered 11 different values ranging from 0 to 1 and separated by steps of 0.1. This range covers the entire spectrum of fractional wettability conditions, including the two most extreme conditions: entirely wettable when $w_{FW} = 1$ and fully water repellent when $w_{FW} = 0$.

According to the criterion proposed by Di Prima et al. (2021), we determined the final time of the infiltration experiment as the greater of either 3$t_{max}$, representing the maximum time for which Eqs. (2) and (4) remain valid, or the time required to recover 95% of the regular infiltration rate, $t_{WR}$. In other words, $t_{WR}$ corresponds to the moment when the quantity $(1 - e^{-\alpha_{WR}t})$ reaches 0.95. Note that the validity time was taken into account for the approximate expansion of the QEI model (Eq. 2) and the approximate expansion (Eq. 4) as defined for the water-repellent case.

We distinguished between transient and steady-state conditions by visually inspecting the linear portion of the infiltration rates versus time plot. The time to steady state, $t_s$ (T), was defined as the time at which the cumulative infiltration curves exhibited approximate linearity with time. This time was considered to represent the conditions under which flow has become primarily gravity driven and all capillarity and water-repellency effects have become negligible. The steady-state infiltration rate, $i_s^{exp}$ (LT$^{-1}$), was estimated by linear regression analysis of cumulative infiltration data after time $t_s$.

### 3.2.2. STEP 2: inversion of synthetic curves

The estimators for $S$, $\alpha_{WR}$ and $w_{FW}$, i.e., $\hat{S}$, $\hat{\alpha}_{WR}$ and $\hat{w}_{FW}$, were obtained by fitting the transient portion of the synthetic data (i.e., data points from time 0 until time $t_s$) to Eq. (5) by minimizing the sum of squared errors (SSE) between synthetic data and modeled cumulative infiltration, $I_{FW}$(T).

Abou Najm et al. (2021) suggested that models modified with the correction term $(1 - e^{-\alpha_{WR}t})$ may face equifinality or other associated uncertainties when solely fitted to infiltration data. In addition, Di Prima et al. (2021) noted that Eq. (4) may lead to an overestimation of $S$ in the case of severe or extreme water repellency. To mitigate such inconveniences when fitting Eq. (5), we constrained the sorptivity to a maximum value, $S_{max}$, determined at the late stage of the transient phase, i.e., when the effect of water repellency has nearly ceased. This procedure involved imposing that the infiltration rate over the time interval $[t_{s-1}, t_s]$ is equal to the experimental steady-state infiltration rate:

$$i_W(t) = i_s^{exp} \qquad \forall t \in [t_{s-1}, t_s] \tag{6}$$

where $i_W(t)$ is determined through Eq. (2) as follows over the time interval $[t_{s-1}, t_s]$:

$$i_W(t) = \frac{I_W(t_s) - I_W(t_{s-1})}{t_s - t_{s-1}} \quad (7)$$

We solved Eq. (6) by identifying the sorptivity value that made the term $i_W(t) - i_s^{exp}$ converge to zero. This procedure assumed that during the steady phase, the soil always experienced wettable conditions, and water infiltration at times immediately preceding this phase can be modeled using Eq. (2), regardless of the actual soil characteristics and curve shapes. For all tests that exhibited evidence of water repellency (e.g., convex curve shapes), this approach required us to determine the sorptivity of the equivalent non-repellent soil, i.e., a soil with the same hydraulic characteristics except for soil water repellency. An example of this process is reported in Figure 2.

The optimization involved 77 sets of initial parameter values for $\hat{\alpha}_{WR}$ and $\hat{w}_{FW}$, with the following starting values: $\hat{\alpha}_{WR}$ = 0.00001, 0.0001, 0.001, 0.01, 0.1, 1 and 10 s$^{-1}$, $\hat{w}_{FW}$ = 0, 0.1, 0.2, 0.3, 0.4, 0.5, 0.6, 0.7, 0.8, 0.9 and 1. In all cases, we used $S_{max}$ as the initial value for soil sorptivity.

The parameter set with the smallest SSE was chosen as the global optimum solution. The estimator for the saturated soil hydraulic conductivity, $\hat{K}_s$, was then estimated as follows (Lassabatere et al., 2006):

$$\hat{K}_s = A\hat{S}^2 - i_s^{exp} \quad (8).$$

The accuracy of these fits was evaluated based on the consistency of the model shape and the fit relative error, $Er_{FIT}$, which was estimated as follows:

$$Er_{FIT} = \sqrt{\frac{\sum_{i=1}^{s-1}[I_{exp}(t_i) - I_{est}(t_i)]^2}{\sum_{i=1}^{s-1} I_{exp}^2(t_i)}} \quad (9)$$

where $(s - 1)$ is the number of data points considered for the transient state and $I_{exp}$ and $I_{est}$ are the experimentally-derived and estimated values for water infiltration.

The inversion procedure was performed using an algorithm coded in R software (R Core Team, 2021), and is accessible on the following webpage https://bestsoilhydro.net/downloads/.

The relative error, $Er$, was also calculated for each estimated value of $\hat{S}$, $\hat{K}_s$, $\hat{\alpha}_{WR}$ and $\hat{w}_{FW}$, and was compared to the corresponding reference value as follows:

$$Er(x) = \frac{\hat{x} - x}{x} \quad (10)$$

where $\hat{x}$ is the estimated value and $x$ is the target, i.e., the reference value $S$, $K_s$, $\alpha_{WR}$ and $w_{FW}$.

## 4. Results and discussion
### 4.1. Analysis of the Berchidda dataset

The Berchidda dataset was categorized into eight shape types based on the criterion proposed by Pachepsky and Karahan (2022), and were ordered based on their abundance as follows (Table 2):

- 24 concave-to-convex-to-linear: similar to **L** type but with a final linear stage;
- 13 convex: **K** type;
- 7 convex-to-linear: **E** type;
- 5 concave-to-linear: **A** shape type;
- 5 slightly concave-to-linear: similar to **A** type but with a less pronounced initial concave part;

- 3 convex-to-concave-to-linear: similar to **F** type but with a more gradual transition stage from the convex to the linear stages;
- 2 slightly concave-to-convex: similar to **L** type but with a less pronounced initial concave part;
- 1 linear: **H** type.

The letters refer to the categories defined by Pachepsky and Karahan (2022) and illustrated in their Figure 2.

The intercept values from the steady-state portion of the curves in Table 2 were identified as being sensitive indicators of the occurrence of water repellency, matching the findings from previous studies (e.g., Loizeau et al., 2017). Specifically, the parameters $w_{FW}$ and $\alpha_{WR}$ both tended to decrease with declining intercept values, corresponding to a change in the curves' shape from concave to convex (Figure 3a and e). This result signals that both parameters are sensitive to the occurrence of water repellency.

The curve shape types had notable differences in the distributions of $w_{FW}$ and $\alpha_{WR}$ values. Median $w_{FW}$ values were ordered as follows (Figure 3b): concave-to-linear, slightly concave-to-linear, concave-to-convex-to-linear, linear, convex-to-concave-to-linear, convex-to-linear, convex, slightly concave-to-convex. For the $\alpha_{WR}$ parameter, the order was as follows (Figure 3f): concave-to-linear, slightly concave-to-linear, linear, convex-to-concave-to-linear, convex-to-linear, convex, concave-to-convex-to-linear, slightly concave-to-convex. Figure 3b and f show almost identical orders with the only exception being the concave-to-convex-to-linear type, which moved from the third position when the curves were organized according to the $w_{FW}$ values to the seventh position when the curves were arranged according to the $\alpha_{WR}$ values. This variation was a consequence of the heterogeneous distribution of the hydrophobic material onto the soil surface, and because the two parameters provide distinct but complementary information. While $\alpha_{WR}$ expresses the rate of attenuation exerted by the repellent fraction on water infiltration, $w_{FW}$ expresses the distribution of this fraction on the soil surface. So, it is possible that a particular soil could exhibits intense impedance phenomena (i.e., low $\alpha_{WR}$) over a substantial portion of the surface. This is the case observed in the response of the concave-to-convex-to-linear curves in Figure 3b and f, where the median $\alpha_{WR}$ value of $2.6\ h^{-1}$ was associated with a median $w_{FW}$ of 0.33 (i.e., 2/3 of the surface exhibited intense impedance phenomena). We argue that an effective interpretation of the bulk behaviour in the infiltration process within a fractional medium is only achievable when both parameters are considered simultaneously. Other fractional wettability scenarios can also include less intensive impedance phenomena that occur on relevant soil fractions, as observed in the response of the convex-to-concave-to-linear curves, which had a median $\alpha_{WR}$ of $76.5\ h^{-1}$ and a median $w_{FW}$ of 0.28.

Regarding concave-shaped curves, it is notable that $w_{FW}$ can assume values of both 0 and 1. As discussed in the Theory section, this phenomenon occurs because Eq. (5) aligns with Eq. (2) under wettable conditions and concave-shaped curves. This condition may be achieved under the following two circumstances: the first and more intuitive, when $w_{FW} = 1$ and $I_{WR}(t) = 0$, and the second, when $w_{FW} = 0$ and $I_{WR}(t)$ assumes the common formulation of Eq. (2) as a consequence of extremely high $\alpha_{WR}$ values (Di Prima et al., 2021). Three curves out of the five that were categorized as concave-to-linear fell within the first scenario and did not exhibit $\alpha_{WR}$ values (Figure 3a and e). Only one fell within the second scenario. For these four curves, the experimental data were modeled by Eq. (2) (Table 2). The last curve was modeled by Eq. (5) and was characterized by an $\alpha_{WR}$ value of $90.9\ h^{-1}$, which represents low impedance phenomena occurring in a soil fraction of 0.71 ($1 - w_{FW}$).

The $w_{FW}$ values exhibited non-normal distributions according to the Kolmogorov-Smirnov test, even after log transformation. Thus, we employed the non-parametric Kruskal-Wallis test to determine whether the median values of this parameter differed between those measured in the open spaces and below tree canopies (Figure 3c-d). According to this test, the $w_{FW}$ median values were significantly different (p-value = 0.001), with lower $w_{FW}$ values below the tree canopies (median $w_{FW} = 0.20$), compared to the open space (median $w_{FW} = 0.32$; Figure 3d). We then compared how the main types of curve shapes were distributed for tests performed in open spaces versus beneath the tree canopy. The majority (73%) of curves exhibiting a convex shape were measured below the tree canopies (16 out of 22), resulting in a median $w_{FW}$ of 0.14. The concave-to-convex-to-linear curves were more evenly distributed, with 42% (10 out of 24) recorded below the tree canopies and the remaining 58% (14 out of 24) in the open space. The $w_{FW}$ values for this shape type were normally distributed; thus, we used a two-sample t-test to determine whether the means of the two groups were different. Although fractional wettability was detected in both locations, the t-test highlighted significant differences, with lower $w_{FW}$ values below the tree canopies (mean $w_{FW} = 0.27$), compared to the open spaces (mean $w_{FW} = 0.36$). These results confirm that locations beneath the tree canopy had more prevalent water-repellent conditions. Specifically, the lower $w_{FW}$ signaled a more homogeneous distribution of the hydrophobic material onto the soil surface of this location.

For each sampled tree and the adjacent open space, we compared the fractional $w_{FW}$ parameter with $WDPT$ measurements. The comparison revealed significant correlations, with a Pearson correlation between $log(median(WDPT))$ and $mean(w_{FW})$ values of –0.948 (p-value = 0.004; Figure 5). The correlation between the two variables can be attributed to the similarity of the provided information. On one hand, $1 - w_{FW}$ expresses the distribution of the water-repellent fraction on the soil surface, enabling meaningful comparisons with other water-repellency-sensitive parameters and indicators. Furthermore, the potential range of values for $w_{FW}$ extends only from 0 to 1, whereas $\alpha_{WR}$ spans values from 0 to +∞. This limited range of values for $w_{FW}$ may enhance comparisons between parameters and among soils with varying responses to water repellency.

### 4.2. Analytical validation
#### 4.2.1. Categorization of the synthetic curves generated with the new model

The generation of synthetic curves through Eq. (1) resulted in a variety of shapes, highlighting the ability of the model to account for diverse hydraulic behaviors. The six soils presented a comparable array of shapes (Appendix 1). As an illustrative example, Figure 5 shows the cumulative infiltration curves for the sandy loam soil. The diverse array of shapes included in the 660 curves were categorized as follows:

- Concave shapes (66 curves), designated as $I_W(t)$. These curves occurred under the following two circumstances: i) when, for each soil, $\alpha_{WR}$ coincided with its maximum value (10000, 5000, 1000, 100, 80, or 10 h$^{-1}$), and ii) when $w_{FW} = 1$, i.e., when the wettable fraction represented 100% of the soil (Figure 1a), regardless of the $\alpha_{WR}$ value;
- Convex shapes (6 curves, one for each soil), designated as $I_{WR}(t)$. These curves occurred when $w_{FW} = 0$, i.e., when the soil surface was simulated as being entirely covered by the water-repellent layer (Figure 1b), and $\alpha_{WR}$ coincided with its minimum value (40, 20, 4, 0.4, 0.32, or 0.04 h$^{-1}$), representing extremely water-repellent conditions;

Mixed shapes (588 curves), designated as $I_{FW}(t)$. These curves included both cases of concave-to-convex and convex-to-concave curves, and occurred under two circumstances: i) when $0 < w_{FW} < 1$, i.e., under fractional wettability conditions (Figure 1c), and ii) when $w_{FW} = 0$ with soil that exhibited intermediate water-repellent conditions expressed by a $α_{WR}$ value between its minimum and maximum.

### 4.2.2. Validation of the inverse procedure

The output parameters of the inversion procedure are reported in Appendix 2. The fitting of Eq. (5) to transient data was accurate, with $Er_{FIT}$ never exceeding 1.2%. In terms of the quality of estimates, the relative error, $Er(α_{WR})$, between the estimated ($\hat{α}_{WR}$) and reference ($α_{WR}$) values consistently decreased with lower $w_{FW}$ values (e.g., Figure 6i, X₁). The least amount of error was associated with $w_{FW} = 0$, in which the entire surface was characterized as being water repellent and the hydrophobic effect on water infiltration has been maximized. On the contrary, errors were maximized when the soil was entirely wettable, with $w_{FW} = 1$ (e.g., Figure 6i, X₂). In other words, when $w_{FW}$ approached zero, $α_{WR}$ became a significant factor in the model, exerting a strong influence. Conversely, as $w_{FW}$ approached one, the model became increasingly insensitive to $α_{WR}$, and its influence diminished. As outlined in the theory section, when $w_{FW} = 1$, the second term of the right-hand side of Eq. (1), $I_{WR}(t)$, equals zero, resulting in infiltration becoming $I_W(t)$ with a concave shape. Consistently, for a regular shape, Eq. (5) yields higher $\hat{α}_{WR}$ values compared to unrealistically low target $α_{WR}$ values. This result arises from the fact that $I_{WR}(t)$ can be used to model regular concave shapes, provided that $α_{WR}$ takes high values.

We also note that $Er(α_{WR})$ increased when the soil was modeled as being more wettable, i.e., having greater $α_{WR}$ values (e.g., Figure 6i, X₃). In this case, the effect of water repellency on water infiltration became attenuated as consequence of the presence of a less-repellent layer covering the soil surface. As a result, the estimation of $\hat{α}_{WR}$ became increasingly uncertain, reaching its apex in the extreme scenario of a completely wettable soil, i.e., when $α_{WR}$ had its maximum value for a given soil (e.g., Figure 6i, X₄). When this circumstance occurs with real data, the soil behavior can be readily discerned by the operator, facilitated by the easily detectable concave shape of the cumulative infiltration and the decreasing trend observed in infiltration rates. In addition, the misestimation of $α_{WR}$ is expected to have negligible impact on the determination of soil sorptivity, particularly when $α_{WR}$ assumes exceptionally high values (Di Prima et al., 2021).

Under water-repellent conditions (indicated by low $α_{WR}$ values) and the presence of a small fraction of water-repellent soils (identified as having relatively high values of $w_{FW}$), the model consistently yielded reliable predictions for $w_{FW}$, indicated by the low $Er(w_{FW})$ values (e.g., Figure 6j, X₅). In contrast, for small $w_{FW}$ values and for attenuated water-repellent conditions. i.e., high $α_{WR}$ values, the differences between the estimator $\hat{w}_{FW}$ and the reference $w_{FW}$ value substantially rose (e.g., Figure 6j, X₆), reaching their peak in the worst-case scenario when $α_{WR}$ assumed its maximum value. This result was because, although the water-repellent fraction increased as $w_{FW}$ approaches 0, the weak effect exerted by water repellency on water infiltration, along with the synthetic curves exhibiting a substantial concave shape, did not allow for a reliable prediction of $w_{FW}$. In other words, the estimations became more accurate when there was a significant contrast between the wettable and water-repellent fractions.

The estimates for $S$ had error values, $Er(S)$, that ranged from −0.8 to 4.7% (Figure 6c, g, k, o, s, w), while $Er(K_s)$ ranged from −8.3 to 1.5% (Figure 6d, h, l, p, t, y). These results represent an improvement compared to previous studies, such as those reported by Di Prima et al. (2021). More broadly, constraining $S$ to its maximum value, $S_{max}$, appears to be a good approach to prevent that parameter from being overestimated. These results also show that Eq. (5) yielded similar outcomes for the entire range of wettability, from entirely wettable to fully water-repellent soils and the various intermediate conditions associated with fractional wettability.

### 4.3. Comparison of representative curves from experimental and synthetic datasets

In this section, we present three comparisons between the experimental and synthetic datasets for representative curves from three distinct shape types: convex, concave, and mixed shapes. Our intent here is to provide a guide for readers to interpret the shape of their infiltration curves and analyze the underlying parameters. In addition, readers may refer to Appendix 3, where the graphical concept of the fractional phenomenon is combined with additional information concerning the $w_{FW}$ and $\alpha_{WR}$ parameters estimated for the Berchidda site. For each comparison, we chose one curve from the Berchidda experimental dataset and then selected a synthetic analogue from the analytical dataset that 1) closely mimicked the experimental curve's shape and 2) was generated using comparable $w_{FW}$ and $\alpha_{WR}$ parameters. We selected all synthetic curves from the sandy loam soil, consistent with the soil texture at the Berchidda site.

For the mixed shape, we selected from the Berchidda dataset the tenth (10) curve sampled below (B) Tree 1 (1) (Table 2; test ID: Berchida_1B_10; Figure 7a). The synthetic analogue was generated by using a $w_{FW}$ value of 0.4 and an $\alpha_{WR}$ value of 4 h$^{-1}$ in Eq. (1) (Figure 7b; test ID: Sandy_Loam_w0.4_alpha4 in Appendix 2). Both curves exhibited mixed shapes characterized by an initial concave part followed by a convex part, with an inflection point, and a final linear part (concave-to-convex-to-linear type). Both curves had relatively small $\alpha_{WR}$ values, indicating intense water repellency phenomena, and $w_{FW}$ values equal to 0.27 for the experimental test and 0.36 for the synthetic test, indicating that the impedance phenomena occurred on 73% to 64% of the infiltration surfaces. Both tests had initial decreases in infiltration rates (Figure 7c-d), which we attributed to the relatively rapid attenuation of the capillarity effect in the wettable fraction. In the second (convex) stage, infiltration rates gradually increased as water impediment decreased in the water-repellent fraction. Once the water-repellency became negligible, the curves approached linearity and gravity becomes the dominant force controlling downward flow through both portions.

For the convex shape, we selected the ninth (9) curve sampled below (B) the canopy of Tree 3 (3) from the Berchidda dataset (Table 2; test ID: Berchida_3B_9; Figure 8a). The synthetic counterpart was generated by using a $w_{FW}$ value of 0 and an $\alpha_{WR}$ value of 4 h$^{-1}$ in Eq. (1) (Figure 8b; test ID: Sandy_Loam_w0_alpha4 in Appendix 2). Both curves displayed the typical convex shape commonly observed under water repellency and were distinguished by low $\alpha_{WR}$ values and $w_{FW} = 0$. It is noteworthy that the water repellency effect was so pronounced that it dominated over any capillarity effect on the ultimate shape of both curves. Indeed, a consistently rising trend of infiltration rates with time was observed in both instances, attributed to the gradual attenuation of the hydrophobic effect as the soil moisture increased (Figure 8c-d). In this scenario, water repellency emerged as the primary factor governing the infiltration process and determining the ultimate shape of the curve.

Finally, for the concave shape, we selected the second (2) curve sampled in the open (O) space adjacent to Tree 2 (2) from the Berchidda dataset (Table 2; test ID: Berchida_2O_2; Figure 9a). The synthetic counterpart was generated by using a $w_{FW}$ value of 0.8 and an $\alpha_{WR}$ value of 100 h$^{-1}$ in Eq. (1) (Figure 9b; test ID: Sandy_Loam_w0.8_alpha100 in Appendix 2). Both curves exhibited the typical concave shape commonly observed under normal conditions, characterized by high $\alpha_{WR}$ values and $w_{FW}$ close to 1. This configuration signifies that infiltration occurred without any water repellency effects anywhere on the surface. The consistent declines in infiltration rates with time seen in these curves were attributable to the gradual attenuation of the capillarity effect as the wetting front advanced deeper into the subsurface. Steady-state conditions were reached between 362 and 2,649 seconds (Figure 9c-d).

The observed similarities between the experimental and synthetic curves support the robustness of the proposed fractional wettability infiltration model. Indeed, the new model successfully reproduces the same types of shapes that can be observed in real soils. These comparisons enhance our ability to accurately interpret infiltration processes and model water infiltration under specific circumstances of fractional wettability. In such situations, the accurate interpretation of the infiltration process necessitates a careful examination of the curve shape, while modeling water infiltration also demands the determination of the extent of the fractional phenomena and knowledge of both the $w_{FW}$ and $α_{WR}$ parameters.

## 5. Summary and conclusions

In this study, we presented a novel infiltration model designed to simulate water infiltration under fractional wettable conditions. This model was tested on the experimental dataset, which comprised 60 cumulative infiltration curves. The interpretation and subsequent categorization of these curves took into account the fractional wettability phenomenon and were based on the observed array of shapes and the values of the model parameters $w_{FW}$ and $α_{WR}$. This assessment also led to the identification of the parameter $w_{FW}$ as a reliable predictor of water repellency occurrence, providing valuable information about the amount of water infiltrating through the two fractions and consequently the fractional distribution of hydrophobic material onto the soil surface. Ad hoc laboratory protocols may be developed in future investigations to assess the physical meaning of the $w_{FW}$ parameter. For instance, soil samples can be repacked, including many known patterns of hydrophobic materials at their surfaces. Simple tests, such as the $WDPT$, can be performed following a gridded scheme in order to characterize the artificial water-repellent layers in terms of persistence of water repellency. Subsequent infiltration tests can be performed to determine $w_{FW}$, which can be compared with the known fractions adopted to repack the samples.

The modeling procedure yielded a set of 660 curves with a diverse range of shapes, encompassing regular concave, convex, and mixed shapes (which included both concave-to-convex and convex-to-concave curves). The model (i.e., Eq. 5) had the ability to account for diverse hydraulic behaviors and depict most realistic infiltration curve types. The subsequent inversion of these curves demonstrated that the model was applicable for interpreting soil parameters such as $S$ and $K_s$ from infiltration tests conducted across various a range of soil wettability conditions. The analysis of the synthetic curves enhanced our ability to recognize specific processes influencing water infiltration, such as water repellency and fractional wettability. The observed similarities between the experimental and synthetic curves demonstrated the robustness of the proposed model. The new model successfully reproduced the same types of shapes observed at the field site, located in Berchidda, Italy. These comparisons further supported the enhanced ability to accurately interpret infiltration processes and model water infiltration under specific circumstances of fractional wettability.

Future investigations may also aim to extend the overarching framework (i.e., Eq. 1) to other infiltration models, including one- and three-dimensional variants and both analytical and numerical solutions. The ability to simulate complex scenarios with various sources of heterogeneity—such as layering, unstable flow, water repellency and fractional wettability—will allow to enhance our understanding of soil water dynamics. To achieve this, numerical simulations would require an ad-hoc adaptation of the Richards equation that integrates the fractional concept and incorporates a new law or new soil hydraulic functions capable of simulating effects analogous to the empirical scaling factor proposed by Abou Najm et al. (2021) for the water-repellent soil fraction.

**Funding**

This work was supported through the project PRIN 2022 PNRR — Methodological proposal for the Individuation of protection forests through LEgislation, geohazard assessment Tools and Ontology (MILETO; project code: P2022587PM), funded by the European Union — Next Generation EU.

**Credit authorship contribution statement**

**Simone Di Prima**: Conceptualization, Methodology, Investigation, Formal analysis, Validation, Visualization, Writing – original draft, Writing - Review & Editing, Funding acquisition. **Ryan D. Stewart**: Writing - Review & Editing. **Majdi R. Abou Najm**: Writing - Review & Editing. **Deniz Yilmaz**: Writing - Review & Editing. **Alessandro Comegna**: Writing - Review & Editing. **Laurent Lassabatere**: Writing - Review & Editing.

**Declaration of Competing Interest**

The authors declare that they have no known competing financial interests or personal relationships that could have appeared to influence the work reported in this paper.

**Declaration of Generative AI and AI-assisted technologies in the writing process**

During the preparation of the first draft the author Simone Di Prima used the tool ChatGPT in order to improve readability and language. After using this tool, all the authors reviewed and edited the content as needed and take full responsibility for the content of the publication.

Table 1. Soil hydraulic parameters and $\alpha_{WR}$ (h$^{-1}$) values for the six studied soils used to model the synthetic cumulative infiltration curves.

| Soil texture | Sand | Loamy Sand | Sandy Loam | Loam | Silt Loam | Silty Clay Loam |
|---|---|---|---|---|---|---|
| $\theta_r$ (m$^3$ m$^{-3}$) | 0.045 | 0.057 | 0.065 | 0.078 | 0.067 | 0.089 |
| $\theta_s$ (m$^3$ m$^{-3}$) | 0.43 | 0.41 | 0.41 | 0.43 | 0.45 | 0.43 |
| $\alpha_{vG}$ (mm$^{-1}$) | 0.0145 | 0.0124 | 0.0075 | 0.0036 | 0.002 | 0.001 |
| $n$ | 2.68 | 2.28 | 1.89 | 1.56 | 1.41 | 1.23 |
| $S$ (mm h$^{-0.5}$) | 86.5 | 58.2 | 36.0 | 20.9 | 16.3 | 6.0 |
| $K_s$ (mm h$^{-1}$) | 297 | 145.9 | 44.2 | 10.44 | 4.5 | 0.7 |
| $l$ | 0.5 | 0.5 | 0.5 | 0.5 | 0.5 | 0.5 |
| $\alpha_{WR}$ (h$^{-1}$) | 10000 | 5000 | 1000 | 100 | 80 | 10 |
|  | 1000 | 500 | 100 | 10 | 8 | 1 |
|  | 800 | 400 | 80 | 8 | 6.4 | 0.8 |
|  | 600 | 300 | 60 | 6 | 4.8 | 0.6 |
|  | 400 | 200 | 40 | 4 | 3.2 | 0.4 |
|  | 200 | 100 | 20 | 2 | 1.6 | 0.2 |
|  | 100 | 50 | 10 | 1 | 0.8 | 0.1 |
|  | 80 | 40 | 8 | 0.8 | 0.64 | 0.08 |
|  | 60 | 30 | 6 | 0.6 | 0.48 | 0.06 |
|  | 40 | 20 | 4 | 0.4 | 0.32 | 0.04 |

Table 2. Berchidda dataset categorization and output parameters.

| Shape type | n. | Test ID | Location | N | n. tr. | n. st. | $i_s^{exp}$ (mm s$^{-1}$) | Intercept (mm) | $w_{FW}$ (−) | $α_{WR}$ (h$^{-1}$) | $S$ (mm h$^{-1}$) | $K_s$ (mm h$^{-1}$) | SSE | Er (%) | Model |
|---|---|---|---|---|---|---|---|---|---|---|---|---|---|---|---|
| Concave-to-convex-to-linear | 1 | Berchidda_1B_1 | below tree canopy | 262 | 157 | 105 | 231.2 | -41.8 | 0.28 | 2.5 | 76.4 | 79.9 | 50.2 | 0.7 | FW |
|  | 2 | Berchidda_1B_3 | below tree canopy | 172 | 93 | 79 | 121.6 | -20.0 | 0.36 | 2.0 | 53.8 | 46.4 | 23.0 | 0.7 | FW |
|  | 3 | Berchidda_1B_4 | below tree canopy | 113 | 34 | 79 | 356.8 | -38.8 | 0.20 | 6.3 | 79.6 | 192.3 | 2.3 | 0.7 | FW |
|  | 4 | Berchidda_1B_6 | below tree canopy | 184 | 107 | 77 | 375.6 | -36.9 | 0.34 | 7.3 | 80.5 | 207.6 | 10.0 | 0.4 | FW |
|  | 5 | Berchidda_1B_8 | below tree canopy | 110 | 45 | 65 | 74.5 | -34.8 | 0.14 | 1.2 | 39.6 | 33.9 | 24.6 | 2.3 | FW |
|  | 6 | Berchidda_1B_9 | below tree canopy | 157 | 71 | 86 | 227.5 | -54.8 | 0.17 | 2.3 | 74.4 | 84.0 | 37.1 | 1.3 | FW |
|  | 7 | Berchidda_1B_10 | below tree canopy | 83 | 47 | 36 | 267.6 | -47.6 | 0.27 | 2.6 | 82.3 | 92.1 | 8.9 | 0.5 | FW |
|  | 8 | Berchidda_1O_1 | open space | 76 | 43 | 33 | 201.2 | -45.8 | 0.25 | 2.9 | 59.8 | 84.3 | 5.1 | 0.5 | FW |
|  | 9 | Berchidda_1O_2 | open space | 140 | 79 | 61 | 111.4 | -78.1 | 0.11 | 1.0 | 48.9 | 33.3 | 16.0 | 0.6 | FW |
|  | 10 | Berchidda_1O_5 | open space | 127 | 37 | 90 | 115.7 | -10.9 | 0.32 | 5.0 | 41.6 | 59.2 | 1.5 | 0.6 | FW |
|  | 11 | Berchidda_1O_7 | open space | 138 | 35 | 103 | 177.2 | -16.8 | 0.32 | 4.6 | 56.4 | 73.5 | 5.5 | 0.9 | FW |
|  | 12 | Berchidda_2B_1 | below tree canopy | 146 | 80 | 66 | 375.9 | -32.4 | 0.33 | 4.8 | 88.4 | 133.3 | 8.6 | 0.5 | FW |
|  | 13 | Berchidda_2B_8 | below tree canopy | 104 | 37 | 67 | 149.1 | -24.6 | 0.35 | 2.4 | 54.8 | 56.0 | 6.1 | 0.7 | FW |
|  | 14 | Berchidda_2O_1 | open space | 119 | 62 | 57 | 490.1 | -8.7 | 0.59 | 13.6 | 87.2 | 241.8 | 13.7 | 0.7 | FW |
|  | 15 | Berchidda_2O_5 | open space | 129 | 72 | 57 | 156.4 | -49.6 | 0.40 | 2.0 | 45.7 | 88.1 | 26.4 | 1.1 | FW |
|  | 16 | Berchidda_2O_8 | open space | 103 | 46 | 57 | 144.1 | -25.0 | 0.36 | 2.2 | 53.1 | 52.1 | 2.6 | 0.4 | FW |
|  | 17 | Berchidda_2O_9 | open space | 105 | 69 | 36 | 334.3 | -33.4 | 0.35 | 8.9 | 69.6 | 176.2 | 17.2 | 0.5 | FW |
|  | 18 | Berchidda_2O_10 | open space | 149 | 87 | 62 | 208.0 | -22.3 | 0.42 | 4.4 | 57.2 | 101.0 | 10.0 | 0.5 | FW |
|  | 19 | Berchidda_3B_6 | below tree canopy | 103 | 58 | 45 | 57.0 | -33.9 | 0.22 | 0.8 | 33.8 | 23.8 | 21.4 | 1.6 | FW |
|  | 20 | Berchidda_3O_2 | open space | 99 | 53 | 46 | 119.3 | -17.9 | 0.46 | 1.8 | 49.4 | 46.6 | 11.6 | 0.8 | FW |
|  | 21 | Berchidda_3O_3 | open space | 121 | 90 | 31 | 107.1 | -70.5 | 0.26 | 0.7 | 50.3 | 31.5 | 39.2 | 0.6 | FW |
|  | 22 | Berchidda_3O_4 | open space | 110 | 76 | 34 | 213.9 | -30.7 | 0.35 | 5.5 | 58.8 | 110.5 | 24.4 | 0.6 | FW |
|  | 23 | Berchidda_3O_5 | open space | 163 | 89 | 74 | 241.4 | -15.9 | 0.62 | 2.9 | 66.7 | 108.4 | 26.1 | 0.7 | FW |
|  | 24 | Berchidda_3O_7 | open space | 127 | 70 | 57 | 146.7 | -57.7 | 0.29 | 1.2 | 57.5 | 48.0 | 18.9 | 0.7 | FW |
| Convex | 1 | Berchidda_1B_5 | below tree canopy | 177 | 82 | 95 | 359.4 | -140.5 | 0.01 | 1.7 | 97.9 | 110.8 | 140.1 | 2.0 | FW |
|  | 2 | Berchidda_2B_2 | below tree canopy | 121 | 67 | 54 | 320.8 | -50.4 | 0.25 | 5.6 | 65.6 | 187.2 | 11.5 | 0.7 | FW |
|  | 3 | Berchidda_2B_4 | below tree canopy | 120 | 51 | 69 | 352.8 | -55.9 | 0.21 | 5.2 | 73.6 | 184.4 | 15.4 | 0.9 | FW |
|  | 4 | Berchidda_2B_5 | below tree canopy | 97 | 44 | 53 | 423.9 | -69.5 | 0.17 | 3.5 | 96.1 | 137.1 | 19.8 | 1.0 | FW |
|  | 5 | Berchidda_2B_6 | below tree canopy | 112 | 74 | 38 | 382.6 | -92.9 | 0.09 | 2.7 | 92.4 | 117.6 | 23.6 | 0.9 | FW |
|  | 6 | Berchidda_2O_4 | open space | 113 | 54 | 59 | 195.2 | -58.2 | 0.29 | 2.4 | 54.9 | 96.7 | 25.5 | 0.9 | FW |
|  | 7 | Berchidda_3B_1 | below tree canopy | 154 | 84 | 70 | 296.0 | -61.1 | 0.35 | 3.5 | 65.9 | 170.1 | 25.3 | 0.8 | FW |
|  | 8 | Berchidda_3B_2 | below tree canopy | 125 | 67 | 58 | 140.6 | -51.7 | 0.21 | 2.7 | 44.8 | 82.3 | 7.5 | 0.5 | FW |
|  | 9 | Berchidda_3B_7 | below tree canopy | 111 | 64 | 47 | 247.2 | -153.1 | 0.13 | 0.7 | 77.5 | 73.0 | 24.2 | 1.1 | FW |
|  | 10 | Berchidda_3B_9 | below tree canopy | 82 | 49 | 33 | 255.9 | -78.8 | 0.03 | 2.5 | 78.3 | 78.3 | 15.2 | 0.7 | FW |
|  | 11 | Berchidda_3B_10 | below tree canopy | 142 | 89 | 53 | 291.9 | -68.2 | 0.14 | 2.7 | 82.2 | 95.8 | 11.8 | 0.5 | FW |
|  | 12 | Berchidda_3O_8 | open space | 85 | 44 | 41 | 248.2 | -83.3 | 0.04 | 2.2 | 74.8 | 81.1 | 4.2 | 0.5 | FW |
|  | 13 | Berchidda_3O_9 | open space | 120 | 66 | 54 | 222.1 | -60.5 | 0.23 | 2.5 | 65.4 | 94.4 | 27.5 | 0.7 | FW |
| Convex-to-linear | 1 | Berchidda_2B_7 | below tree canopy | 119 | 44 | 75 | 195.4 | -51.7 | 0.12 | 2.5 | 63.4 | 70.6 | 6.9 | 0.8 | FW |
|  | 2 | Berchidda_2B_9 | below tree canopy | 128 | 50 | 78 | 218.6 | -15.9 | 0.10 | 10.0 | 58.3 | 113.0 | 2.3 | 0.7 | FW |
|  | 3 | Berchidda_2B_10 | below tree canopy | 153 | 94 | 59 | 151.6 | -34.7 | 0.26 | 3.2 | 51.1 | 70.7 | 21.3 | 0.6 | FW |
|  | 4 | Berchidda_2O_3 | open space | 215 | 49 | 166 | 154.2 | -28.9 | 0.30 | 4.7 | 42.1 | 96.2 | 3.8 | 0.7 | FW |
|  | 5 | Berchidda_2O_6 | open space | 107 | 60 | 47 | 152.1 | -29.8 | 0.26 | 3.0 | 51.8 | 64.5 | 8.6 | 0.6 | FW |
|  | 6 | Berchidda_3B_4 | below tree canopy | 94 | 46 | 48 | 47.7 | -51.8 | 0.15 | 0.6 | 32.8 | 16.6 | 6.9 | 0.6 | FW |
|  | 7 | Berchidda_3B_5 | below tree canopy | 131 | 100 | 31 | 245.8 | -46.6 | 0.19 | 3.7 | 70.7 | 101.0 | 12.1 | 0.4 | FW |
| Concave-to-linear | 1 | Berchidda_1B_7 | below tree canopy | 187 | 85 | 102 | 250.2 | 19.6 | 0.00 | 855.6 | 71.6 | 117.3 | 30.2 | 1.0 | W |
|  | 2 | Berchidda_1O_9 | open space | 138 | 44 | 94 | 184.6 | 22.6 | 1.00 | NA | 57.9 | 75.1 | 191.7 | 4.7 | W |
|  | 3 | Berchidda_1O_10 | open space | 180 | 101 | 79 | 223.7 | 29.3 | 1.00 | NA | 66.4 | 79.6 | 1129.7 | 5.0 | W |
|  | 4 | Berchidda_2B_3 | below tree canopy | 138 | 69 | 69 | 426.9 | 17.2 | 0.29 | 90.9 | 90.1 | 175.0 | 87.8 | 1.9 | FW |
|  | 5 | Berchidda_2O_7 | open space | 136 | 58 | 78 | 416.4 | 25.9 | 1.00 | NA | 89.0 | 158.0 | 281.3 | 3.4 | W |
| Slightly concave-to-linear | 1 | Berchidda_1B_2 | below tree canopy | 117 | 17 | 100 | 93.3 | 2.1 | 0.77 | 10.3 | 32.1 | 66.7 | 2.3 | 2.6 | FW |
|  | 2 | Berchidda_1O_8 | open space | 164 | 22 | 142 | 238.6 | 1.7 | 0.21 | 246.3 | 46.4 | 168.2 | 0.5 | 0.8 | FW |
|  | 3 | Berchidda_2O_2 | open space | 97 | 19 | 78 | 266.9 | 3.2 | 0.69 | 84.3 | 51.8 | 179.3 | 0.4 | 0.9 | FW |
|  | 4 | Berchidda_3B_8 | below tree canopy | 105 | 8 | 97 | 234.5 | 0.0 | 0.90 | 3600.0 | 32.7 | 203.6 | 0.1 | 1.1 | FW |
|  | 5 | Berchidda_3O_1 | open space | 70 | 8 | 62 | 59.3 | 3.3 | 1.00 | NA | 19.4 | 48.1 | 0.3 | 2.2 | W |
| Convex-to-concave-to-linear | 1 | Berchidda_1O_3 | open space | 138 | 67 | 71 | 665.0 | -6.9 | 0.28 | 76.5 | 93.4 | 380.4 | 8.0 | 0.7 | FW |
|  | 2 | Berchidda_1O_4 | open space | 121 | 60 | 61 | 789.0 | 17.9 | 0.07 | 280.2 | 120.5 | 315.1 | 121.6 | 2.1 | FW |
|  | 3 | Berchidda_1O_6 | open space | 99 | 48 | 51 | 243.2 | 14.1 | 0.38 | 26.2 | 67.9 | 92.7 | 13.1 | 0.7 | FW |
| Slightly concave-to-convex | 1 | Berchidda_3B_3 | below tree canopy | 339 | 298 | 41 | 117.3 | -154.1 | 0.00 | 0.6 | 55.2 | 29.1 | 1089.0 | 1.8 | FW |
|  | 2 | Berchidda_3O_6 | open space | 133 | 115 | 18 | 114.3 | -99.7 | 0.15 | 0.7 | 53.0 | 30.4 | 471.8 | 1.8 | FW |
| Linear | 1 | Berchidda_3O_10 | open space | 150 | 13 | 137 | 569.7 | 0.4 | 0.29 | 95.1 | 78.8 | 384.3 | 0.4 | 1.3 | FW |

Figure 1. Concept of the fractional wettability approach.

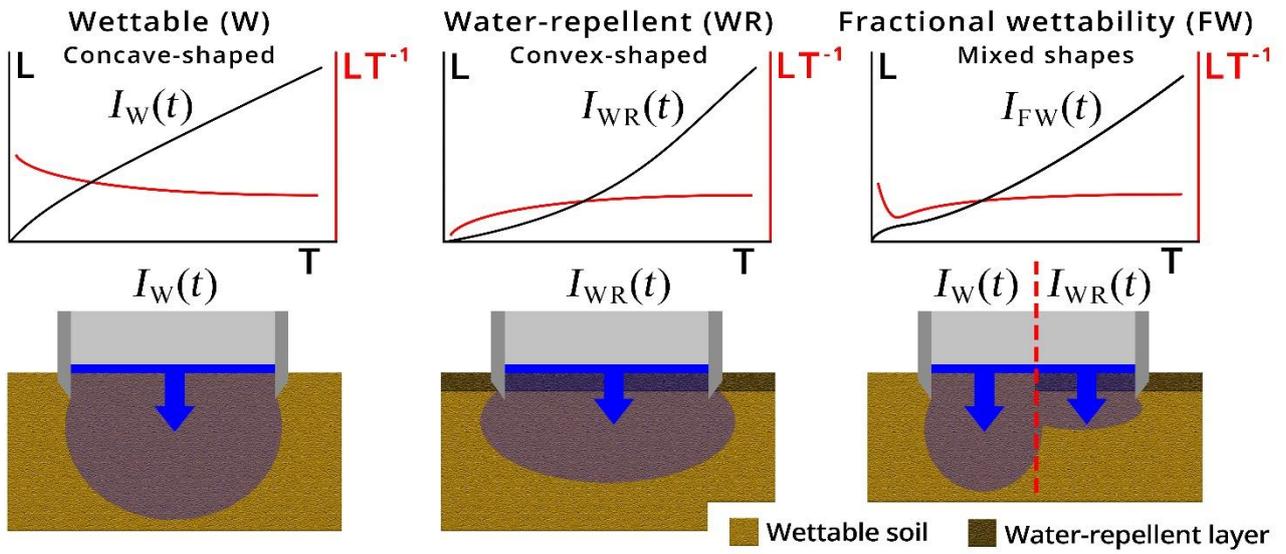

Figure 2. Procedure for estimating the maximum sorptivity, $S_{max}$.

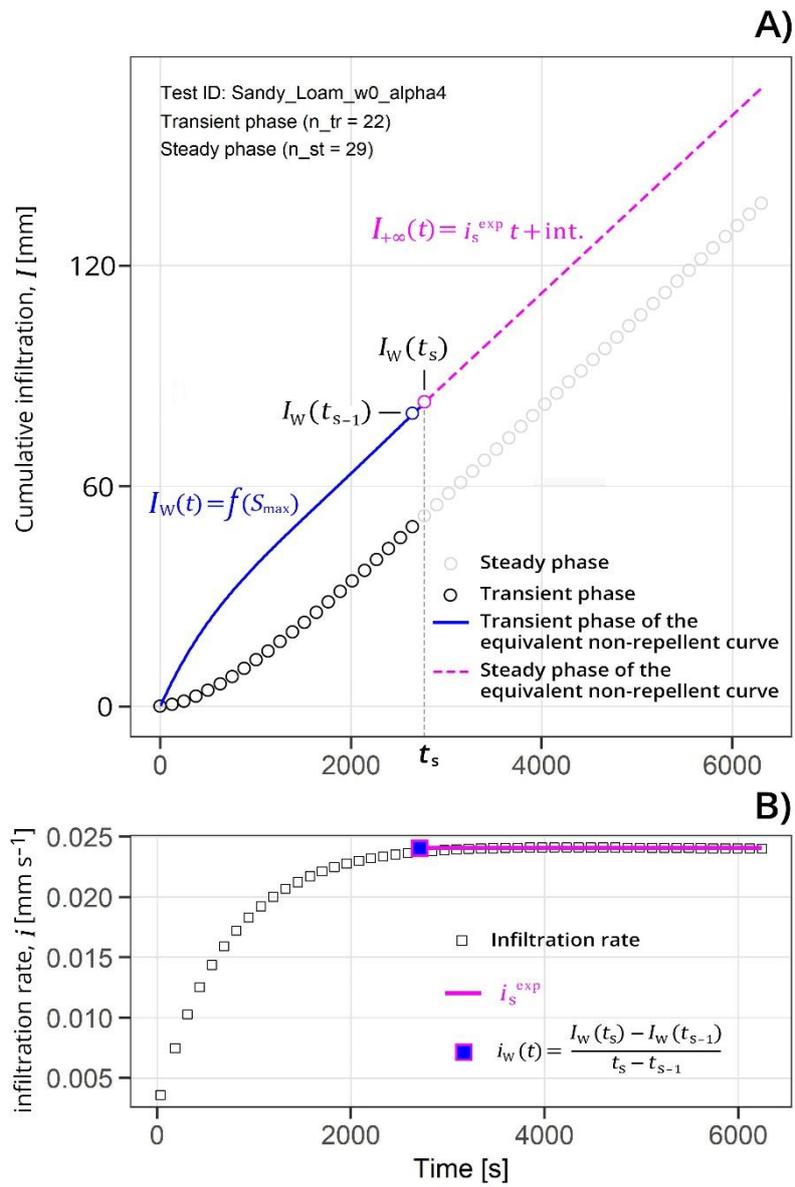

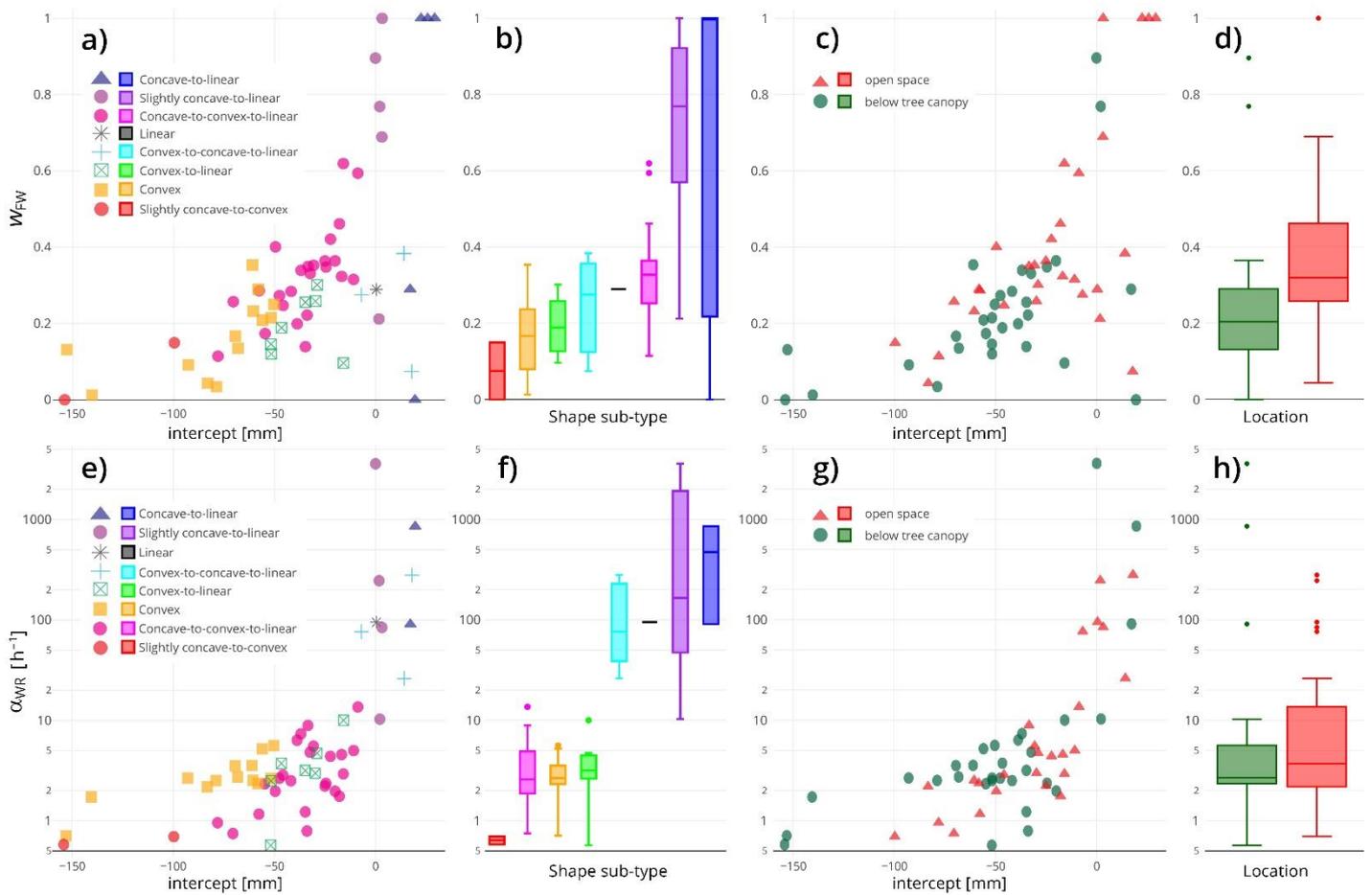

Figure 3. Plots of the $w_{FW}$ (a and c) and $\alpha_{WR}$ (e and g) parameters versus the intercept values obtained from the steady-state portion of the curves. Box plots of the $w_{FW}$ (b and d) and $\alpha_{WR}$ (f and h) parameters. The values are categorized according to the shape types (a, b, e, f) and locations (c, d, g, h).

Figure 4. Plot of representative values of *WDPT* (s) against $w_{FW}$ for three trees.

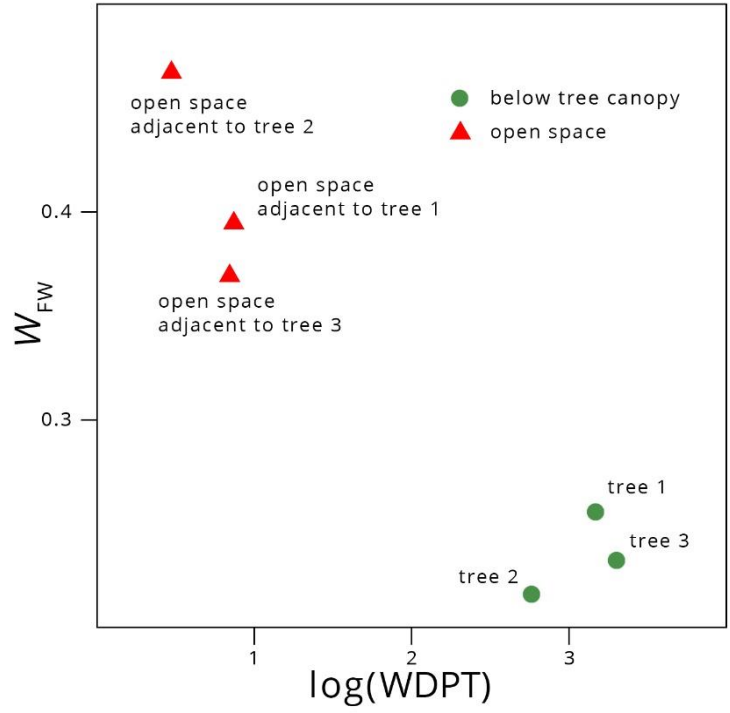

Figure 5. Cumulative infiltration curves for the Sandy Loam soil for different α$_{WR}$ and w$_{FW}$ values. The curves for all the six studied soils were reported in Appendix 1. The curves were generated analytically using Eq. (1) and the Carsel and Parrish (1988) hydraulic parameters (Table 1).

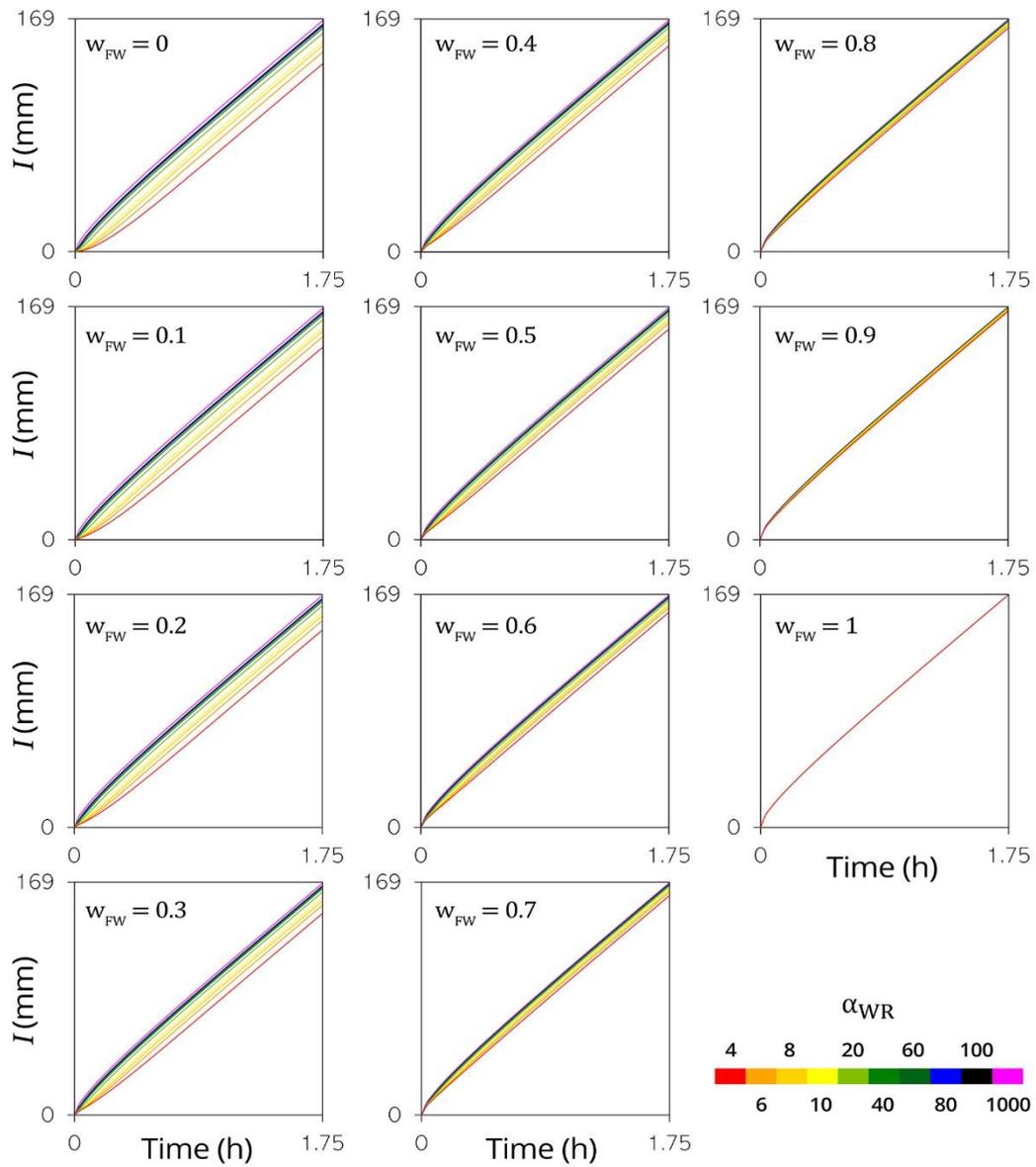

Figure 6. Relative error of the output parameters for the six synthetic soils.

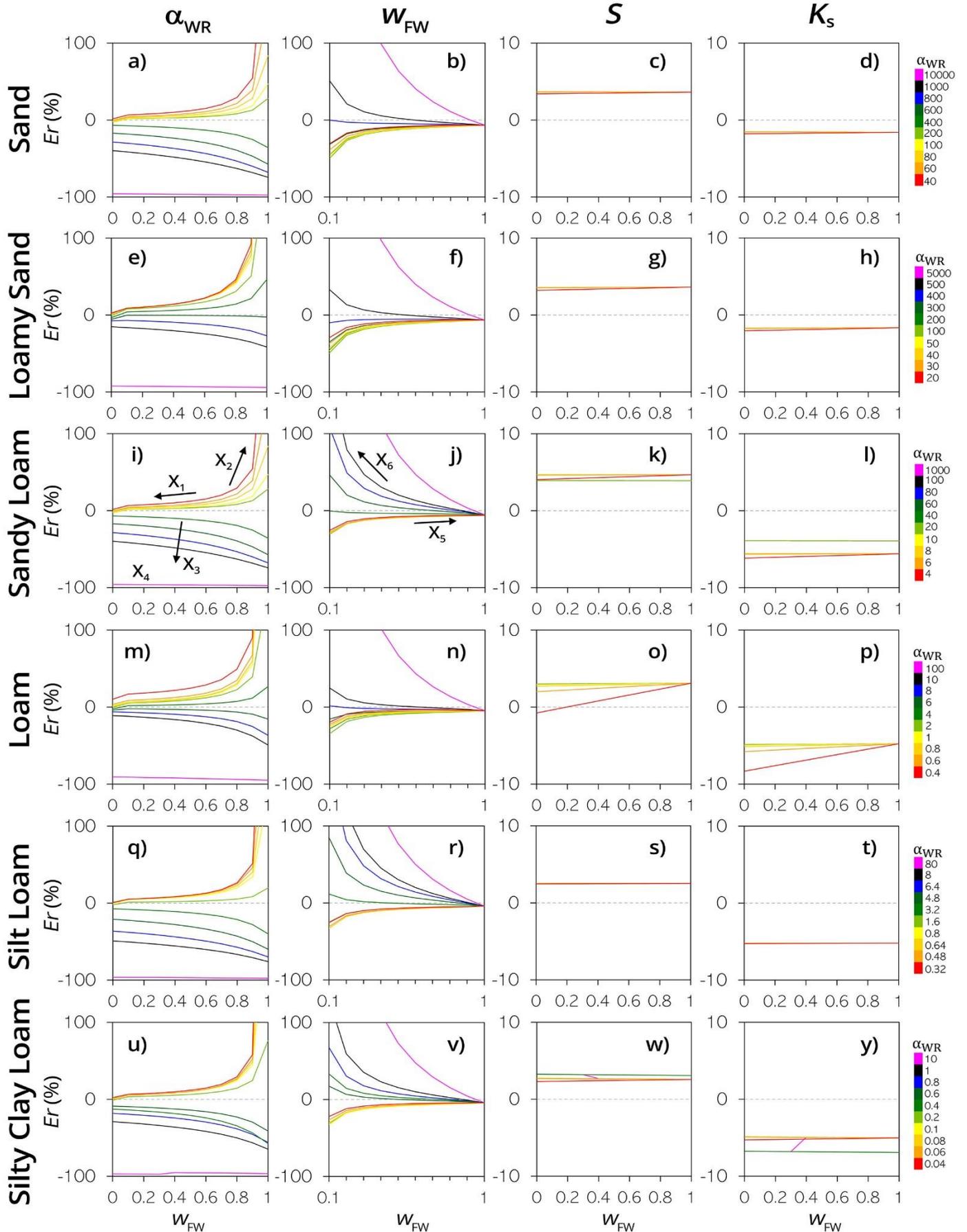

Figure 7. Comparison of representative concave-to-convex-to-linear shaped curves from in-field (left) and synthetic (right) infiltration tests.

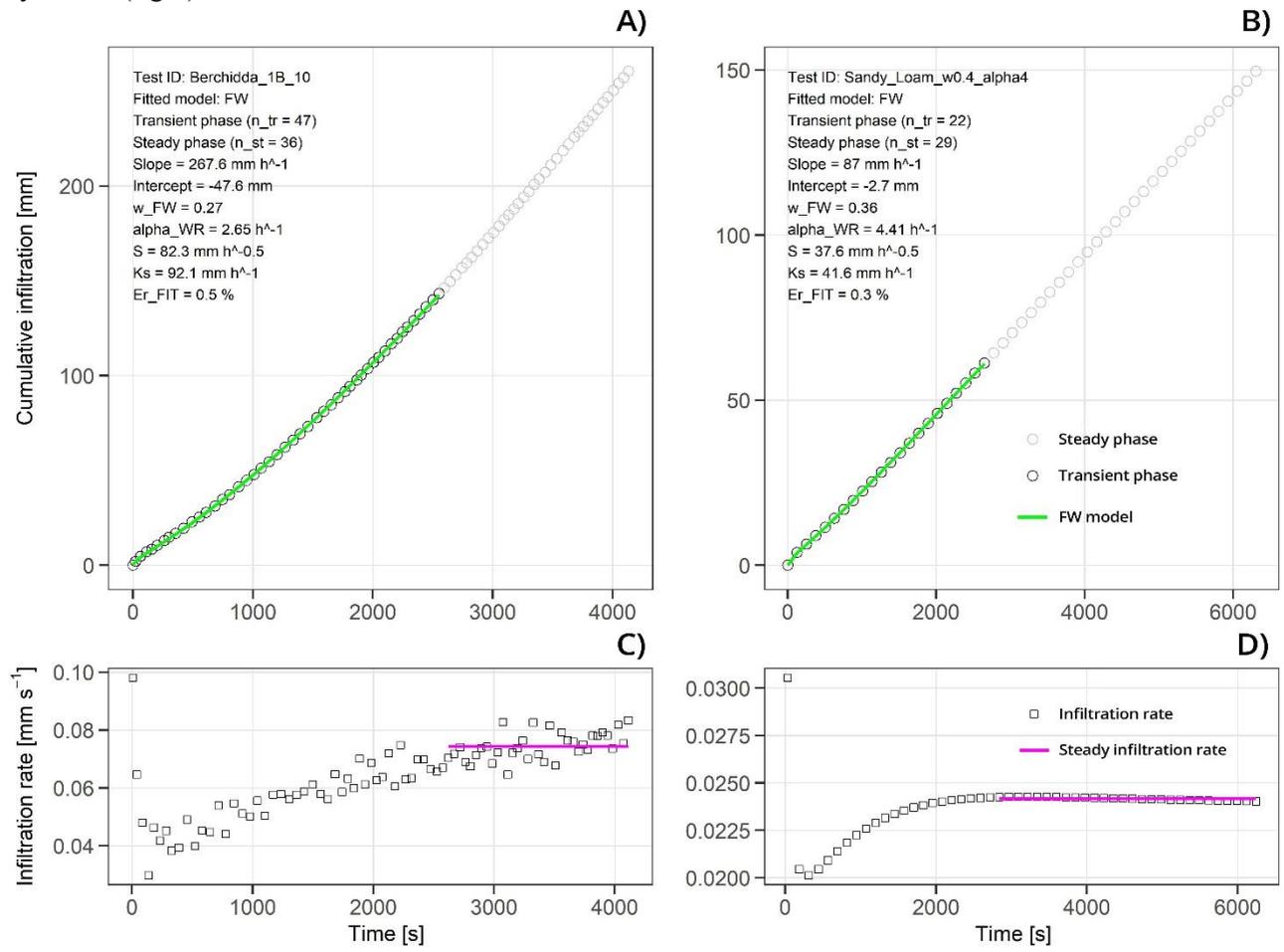

Figure 8. Comparison of representative convex shaped curves from in-field (left) and synthetic (right) infiltration tests.

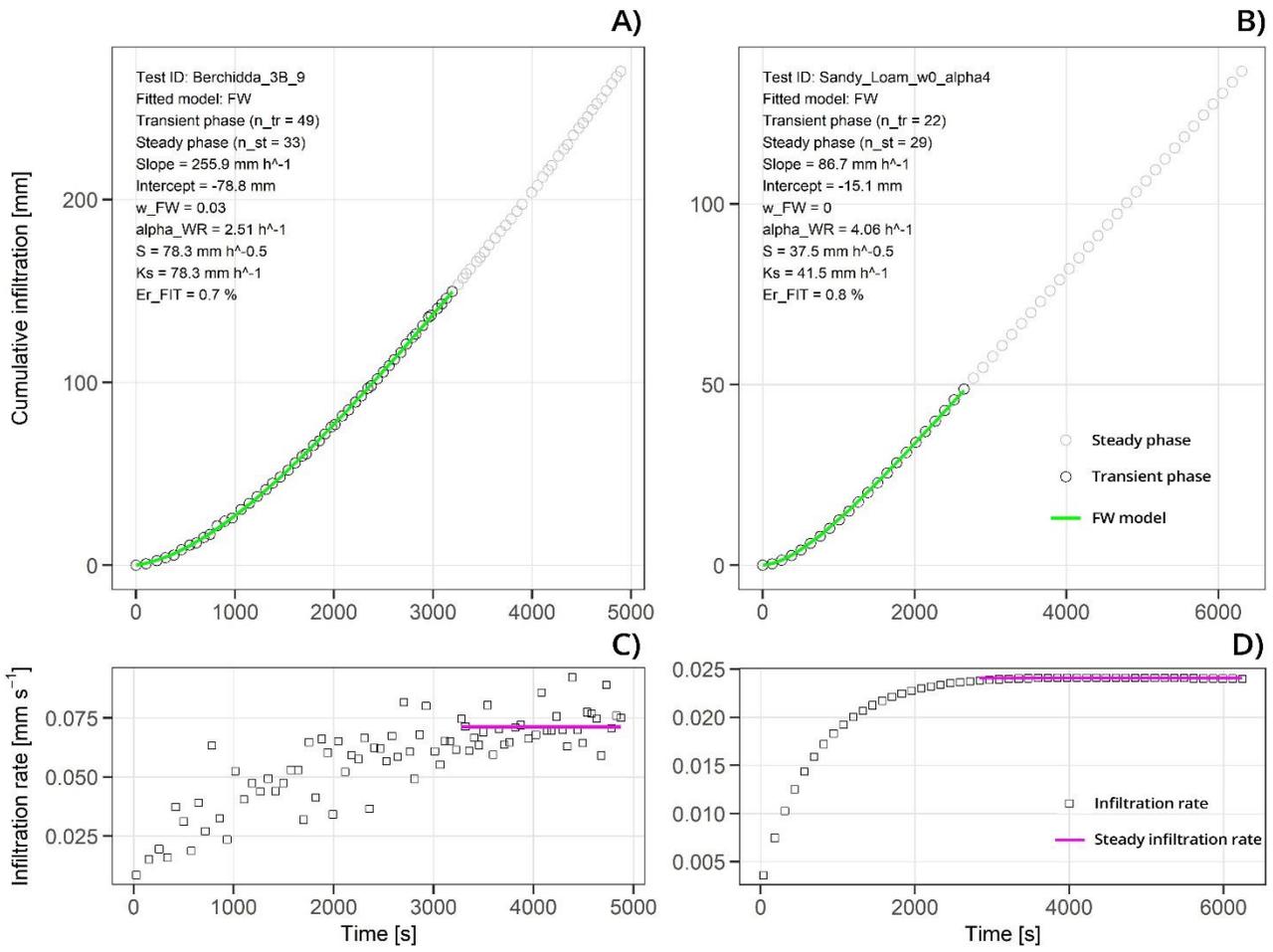

Figure 9. Comparison of representative concave shaped curves from in-field (left) and synthetic (right) infiltration tests.

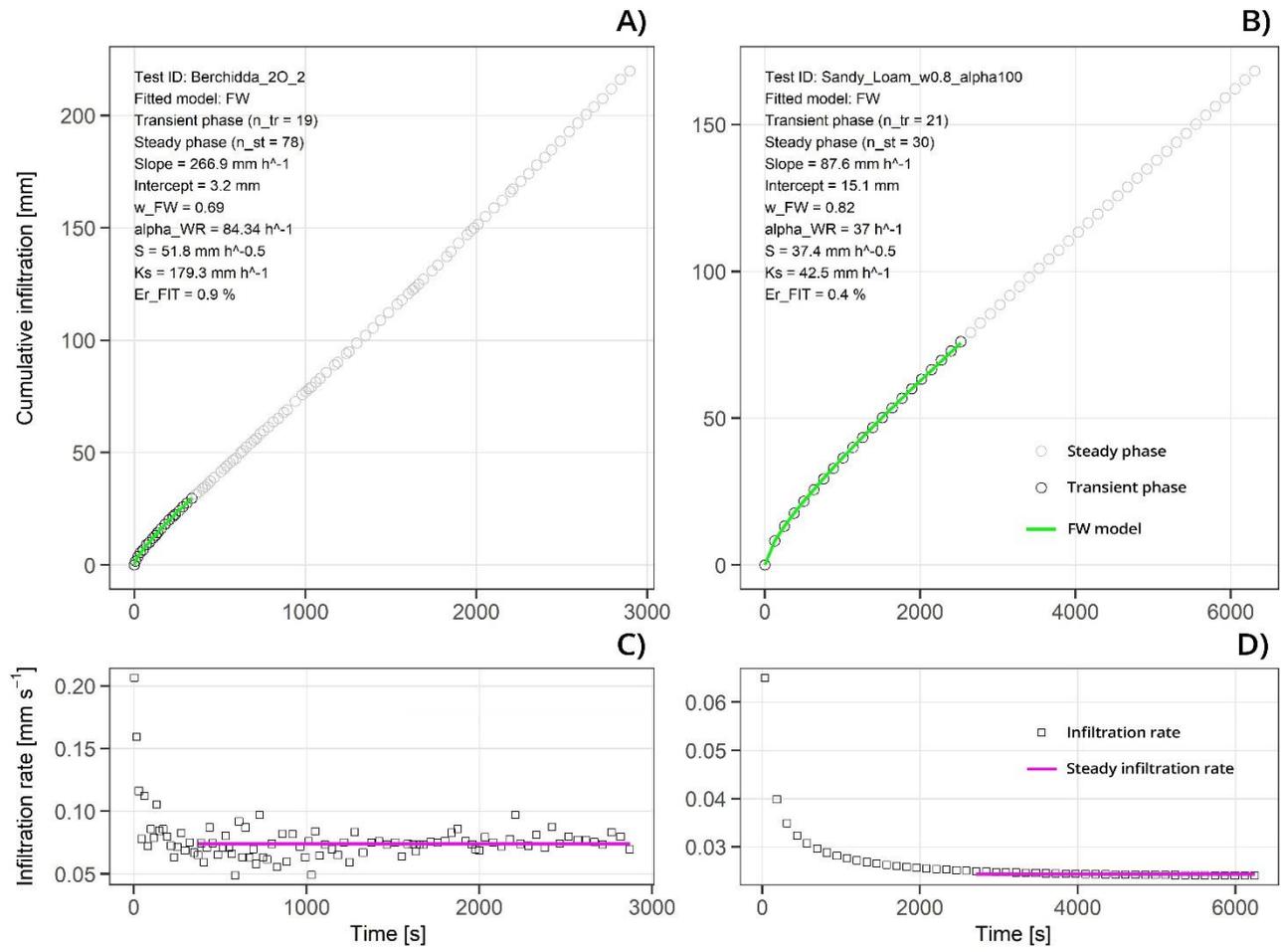